\documentclass[11pt]{article}
\usepackage{fullpage}
\usepackage[fleqn]{amsmath}

\newcommand{\beq}{
   \vspace{-2mm}
   \begin{equation} 
}
\newcommand{\eeq}{
   \end{equation} 
   \vspace{-2mm}
}

\begin{document}

\begin{center} 
{\LARGE \bf Hadoop Performance Models} \\
\vspace{8mm}
{\Large Herodotos Herodotou \\
{\tt hero@cs.duke.edu} \\
\vspace{5mm}
Technical Report, CS-2011-05 \\
Computer Science Department \\
Duke University
}
\vspace{15mm}
\end{center} 

\begin{abstract}
Hadoop MapReduce is now a popular choice for performing
large-scale data analytics. This technical report describes
a detailed set of mathematical performance models for
describing the execution of a MapReduce job on Hadoop.
The models describe dataflow and cost information at the
fine granularity of phases within the map and reduce tasks
of a job execution. The models can be used to estimate the
performance of MapReduce jobs as well as to find the
optimal configuration settings to use when running the jobs.
\end{abstract}

\noindent The execution of a MapReduce job is broken down into 
map tasks and reduce tasks. Subsequently, map task execution is 
divided into the phases: {\em Read} (reading map inputs),
{\em Map} (map function processing),
{\em Collect} (serializing to buffer and partitioning),
{\em Spill} (sorting, combining, 
compressing, and writing map outputs to local disk), 
and {\em Merge} (merging sorted spill files).
Reduce task execution is divided into the phases:
{\em Shuffle} (transferring map outputs to reduce tasks, with
decompression if needed), 
{\em Merge} (merging sorted map outputs), 
{\em Reduce} (reduce function processing), and 
{\em Write} (writing reduce outputs to the distributed 
file-system).
Each phase represents an important
part of the job's overall execution in Hadoop.
We have developed performance models for each task phase,
which are then combined to form the overall Map-Reduce Job model.

\section{Model Parameters}
\label{sec:parameters}

\noindent The performance models rely on a set of parameters to estimate
the cost of a Map-Reduce job. We separate the parameters into three
categories:

\begin{enumerate}
{
\item {\em Hadoop Parameters}: A set of Hadoop-defined configuration
parameters that effect the execution of a job

\item {\em Profile Statistics}: A set of statistics specifying properties
of the input data and the user-defined functions (Map, Reduce, Combine)

\item {\em Profile Cost Factors}: A set of parameters that define the I/O,
CPU, and network cost of a job execution
}
\end{enumerate}

\noindent Table \ref{table:hadoop-params} defines the variables that
are associated with Hadoop parameters. \\

\begin{table}[ht]
{\small
	\centering
	\begin{tabular}{| l | l | c | c |}
	\hline
	Variable & Hadoop Parameter & Default Value	& Effect \\
	\hline
	\hline
	pNumNodes		&	Number of Nodes						&	&	System	\\
	pTaskMem		&	mapred.child.java.opts				&	-Xmx200m	&	System	\\
	pMaxMapsPerNode	&	mapred.tasktracker.map.tasks.max	&	2	&	System	\\
	pMaxRedPerNode	&	mapred.tasktracker.reduce.tasks.max	&	2	&	System	\\
	\hline
	pNumMappers		&	mapred.map.tasks					&	&	Job	\\
	pSortMB			&	io.sort.mb							&	100 MB	&	Job	\\
	pSpillPerc		&	io.sort.spill.percent				&	0.8	&	Job	\\
	pSortRecPerc	&	io.sort.record.percent				&	0.05	&	Job	\\
	pSortFactor		&	io.sort.factor						&	10	&	Job	\\
	pNumSpillsForComb	&	min.num.spills.for.combine		&	3	&	Job	\\
	\hline
	pNumReducers		&	mapred.reduce.tasks				&	&	Job	\\
	pInMemMergeThr		&	mapred.inmem.merge.threshold	&	1000	&	Job	\\
	pShuffleInBufPerc	&	mapred.job.shuffle.input.buffer.percent	&	0.7	&	Job	\\
	pShuffleMergePerc 	&	mapred.job.shuffle.merge.percent		&	0.66	&	Job	\\
	pReducerInBufPerc	&	mapred.job.reduce.input.buffer.percent	&	0	&	Job	\\
	\hline
	pUseCombine	&	mapred.combine.class or mapreduce.combine.class	&	null	&	Job	\\
	pIsIntermCompressed	&	mapred.compress.map.output			&	false	&	Job	\\
	pIsOutCompressed	&	mapred.output.compress					&	false	&	Job	\\
	\hline
	pReduceSlowstart	&	mapred.reduce.slowstart.completed.maps 	&	0.05	&	Job	\\
	\hline
	pIsInCompressed		&	Whether the input is compressed or not 	&	&	Input	\\
	pSplitSize			&	The size of the input split				&	&	Input	\\
	\hline
	\hline
	\end{tabular}
	\caption{Variables for Hadoop Parameters}
    \label{table:hadoop-params}
}
\end{table}

\noindent Table \ref{table:prof-stats} defines the necessary 
profile statistics specific to a job and the data it is processing. \\

\begin{table}[ht]
{\small
	\centering
	\begin{tabular}{| l | p{4.8in} |}
	\hline
	Variable & Description	\\
	\hline
	\hline
	sInputPairWidth	&	The average width of the input K-V pairs	\\
	sMapSizeSel		&	The selectivity of the mapper in terms of size	\\
	sMapPairsSel	&	The selectivity of the mapper in terms of number of K-V pairs	\\
	sReduceSizeSel	&	The selectivity of the reducer in terms of size \\
	sReducePairsSel	&	The selectivity of the reducer in terms of number of K-V pairs \\
	sCombineSizeSel	&	The selectivity of the combine function in terms of size \\
	sCombinePairsSel	&	The selectivity of the combine function in number of K-V pairs \\
	\hline
	sInputCompressRatio	&	The ratio of compression for the input data \\
	sIntermCompressRatio &	The ratio of compression for the intermediate map output \\
	sOutCompressRatio	&	The ratio of compression for the final output of the job \\
	\hline
	\hline
	\end{tabular}
	\caption{Variables for Profile Statistics}
    \label{table:prof-stats}
}
\end{table}

\noindent Table \ref{table:prof-costs} 
defines system specific parameters needed
for calculating I/O, CPU, and network costs.
The IO costs and CPU costs related to compression
are defined in terms of time per byte.
The rest CPU costs are defined in terms of time per K-V pair.
The network cost is defined in terms of transferring time per byte. \\

\begin{table}[ht]
{\small
	\centering
	\begin{tabular}{| l | l |}
	\hline
	Variable & Description	\\
	\hline
	\hline
	cHdfsReadCost			&	The cost for reading from HDFS	\\
	cHdfsWriteCost			&	The cost for writing to HDFS	\\
	cLocalIOCost			&	The cost for performing I/O from the local disk	\\
	cNetworkCost			&	The network transferring cost \\
	\hline
	cMapCPUCost				&	The CPU cost for executing the map function \\
	cReduceCPUCost			&	The CPU cost for executing the reduce function \\
	cCombineCPUCost			&	The CPU cost for executing the combine function \\
	\hline
	cPartitionCPUCost		&	The CPU cost for partitioning \\
	cSerdeCPUCost			&	The CPU cost for serialization \\
	cSortCPUCost			&	The CPU cost for sorting on keys \\
	cMergeCPUCost			&	The CPU cost for merging \\
	\hline
	cInUncomprCPUCost		&	The CPU cost for uncompressing the input data \\
	cIntermUncomprCPUCost 	&	The CPU cost for uncompressing the intermediate data \\
	cIntermComprCPUCost		&	The CPU cost for compressing the intermediate data \\
	cOutComprCPUCost		&	The CPU cost for compressing the output data \\
	\hline
	\hline
	\end{tabular}
	\caption{Variables for Profile Cost Factors}
    \label{table:prof-costs}
}
\end{table}

\noindent Let's define the identity function $I$ as:
\beq\mathit{
I(x) = 
  \begin{cases} 
   1 & \text{if x exists or equals true} \\
   0 & \text{otherwise }
  \end{cases}
}\eeq

\noindent {\bf Initializations:} In an effort present concise formulas
and avoid the use of conditionals as much as possible, we make the
following initializations:

\begin{align}
	& \text{If } (\mathit{pUseCombine} == \text{FALSE}) \nonumber \\
	& \ \ \ \mathit{sCombineSizeSel} = 1 \nonumber \\
	& \ \ \ \mathit{sCombinePairsSel} = 1 \nonumber \\
	& \ \ \ \mathit{cCombineCPUCost} = 0 \nonumber
\end{align}
\begin{align}
	& \text{If } (\mathit{pIsInCompressed} == \text{FALSE}) \nonumber \\
	& \ \ \ \mathit{sInputCompressRatio} = 1 \nonumber \\
	& \ \ \ \mathit{cInUncomprCPUCost} = 0 \nonumber
\end{align}
\begin{align}
	& \text{If } (\mathit{pIsIntermCompressed} == \text{FALSE}) \nonumber \\
	& \ \ \ \mathit{sIntermCompressRatio} = 1 \nonumber \\
	& \ \ \ \mathit{cIntermUncomprCPUCost} = 0 \nonumber \\
	& \ \ \ \mathit{cIntermComprCPUCost} = 0 \nonumber
\end{align}
\begin{align}
	& \text{If } (\mathit{pIsOutCompressed} == \text{FALSE}) \nonumber \\
	& \ \ \ \mathit{sOutCompressRatio} = 1 \nonumber \\
	& \ \ \ \mathit{cOutComprCPUCost} = 0 \nonumber
\end{align}

\section{Performance Models for the Map Task Phases}
\label{sec:map-phases}

\noindent The Map Task execution is divided into five phases:
\begin{enumerate}
{
\item {\em Read}: Reading the input split and creating the key-value pairs.

\item {\em Map}: Executing the user-provided map function.

\item {\em Collect}: Collecting the map output into a buffer and partitioning.

\item {\em Spill}: Sorting, using the combiner if any, performing compression 
if asked, and finally spilling to disk, creating {\em file spills}.

\item {\em Merge}: Merging the file spills into a single map output file.
Merging might be performed in multiple rounds.
}
\end{enumerate}

\subsection{Modeling the Read and Map Phases}
\label{sec:map-read-map-phase}

\noindent During this phase, the input split is read, uncompressed if necessary,
the key-value pairs are created, and passed an input to the user-defined map function. 

\beq\mathit{ inputMapSize = \frac{pSplitSize}{sInputCompressRatio} }\eeq

\beq\mathit{ inputMapPairs = \frac{inputMapSize}{sInputPairWidth} }\eeq

The costs of this phase are:
\begin{align}
\mathit{IOCost_{Read} = pSplitSize \times cHdfsReadCost} \nonumber 
\end{align}
\begin{align}
\mathit{CPUCost_{Read}} &= \mathit{pSplitSize \times cInUncomprCPUCost} \nonumber \\
	& {\ \ \ } + \mathit{inputMapPairs \times cMapCPUCost}
\end{align}

\noindent If the MR job consists only of mappers (i.e. $\mathit{pNumReducers = 0}$), then the 
spilling and merging phases will not be executed and the map output
will be written directly to HDFS.

\beq\mathit{ outMapSize = inputMapSize \times sMapSizeSel }\eeq
\beq\mathit{ IOCost_{MapWrite} = outMapSize \times {sOutCompressRatio} \times cHdfsWriteCost }\eeq
\beq\mathit{ CPUCost_{MapWrite} = outMapSize \times cOutComprCPUCost }\eeq

\subsection{Modeling the Collect and Spill Phases}
\label{sec:map-collect-spill-phase}

\noindent The map function generates output key-value (K-V) pairs that 
are placed in the map-side memory buffer.
The formulas regarding the map output are:

\beq\mathit{ outMapSize = inputMapSize \times sMapSizeSel }\eeq
\beq\mathit{ outMapPairs = inputMapPairs \times sMapPairsSel }\eeq
\beq\mathit{ outPairWidth = \frac{outMapSize}{outMapPairs} }\eeq

\noindent The memory buffer is split into two parts: the 
{\em serialization} part that stores the key-value pairs, 
and the {\em accounting} part that stores metadata per pair. 
When either of these two parts fills up 
(based on the threshold value $\mathit{pSpillPerc}$),
the pairs are partitioned, sorted, and spilled to disk.

\noindent The maximum number of pairs for the serialization buffer is:

\beq\mathit{ maxSerPairs = \left\lfloor 
     \frac{pSortMB \times 2^{20} \times (1 - pSortRecPerc) \times pSpillPerc}
          {outPairWidth} \right\rfloor }\eeq

\noindent The maximum number of pairs for the accounting buffer is:

\beq\mathit{ maxAccPairs = \left\lfloor 
     \frac{pSortMB \times 2^{20} \times pSortRecPerc \times pSpillPerc}
          {16} \right\rfloor }\eeq

\noindent Hence, the number of pairs and size of the buffer before a spill will be:

\beq\mathit{ spillBufferPairs = Min\{\ maxSerPairs,\ maxAccPairs\ ,\ outMapPairs\ \} }\eeq
\beq\mathit{ spillBufferSize = spillBufferPairs \times outPairWidth }\eeq

\noindent The overall number of spills will be:

\beq\mathit{ numSpills = \left\lceil \frac{outMapPairs}{spillBufferPairs} \right\rceil }\eeq

\noindent The number of pairs and size of each spill 
depends on the width of each K-V pair,
the use of the combine function, and the use of intermediate data compression.
Note that $\mathit{sIntermCompressRatio}$ is set to $1$ by default, 
if intermediate compression is disabled.
Note that $\mathit{sCombinePairsSel}$ and $\mathit{sCombinePairsSel}$ are set to $1$ by default, 
if no combine function is used.

\beq\mathit{ spillFilePairs = spillBufferPairs \times sCombinePairsSel }\eeq
\beq\mathit{ spillFileSize = spillBufferSize \times sCombineSizeSel \times sIntermCompressRatio }\eeq

\noindent  The costs of this phase are:
\begin{align}
\mathit{IOCost_{Spill} = numSpills \times spillFileSize \times cLocalIOCost}
\end{align}
\begin{align}
CPUCost_{Spill} &= \mathit{numSpills} \ \times \nonumber \\
	&\hspace{-10mm} [\ \mathit{spillBufferPairs \times cPartitionCPUCost} \nonumber \\
	&\hspace{-10mm} + \mathit{spillBufferPairs \times cSerdeCPUCost} \nonumber \\
    &\hspace{-10mm} + \mathit{spillBufferPairs \times 
			\log_2 (\frac{spillBufferPairs}{pNumReducers}) \times cSortCPUCost} \nonumber \\
    &\hspace{-10mm} + \mathit{spillBufferPairs \times cCombineCPUCost} \nonumber \\
    &\hspace{-10mm} + \mathit{spillBufferSize \times sCombineSizeSel \times cIntermComprCPUCost}\ ]
\end{align}

\subsection{Modeling the Merge Phase}
\label{sec:map-merge-phase}

\noindent The goal of the merge phase is to merge all the spill files into 
a single output file, which is written to local disk. The merge phase will
occur only if more that one spill file is created. Multiple merge passes
might occur, depending on the {\em pSortFactor} parameter.
We define a {\em merge pass} to be the merging of at most $\mathit{pSortFactor}$ spill files.
We define a {\em merge round} to be one or more merge passes that merge only spills
produced by the spill phase or a previous merge round. 
For example, suppose $\mathit{numSpills = 30}$ and $\mathit{pSortFactor = 10}$. 
Then, $3$ merge passes will be performed to create $3$ new files.
This is the first merge round. Then, the $3$ new files will be merged
together forming the 2nd and final merge round. \\

\noindent The final merge pass is unique in the sense that if the number
of spills to be merged is greater than or equal to $\mathit{pNumSpillsForComb}$, 
the combiner will be used again. Hence, we treat the intermediate merge
rounds and the final merge separately. For the intermediate merge passes,
we calculate how many times (on average) a single spill will be read. \\

\noindent Note that the remaining section assumes $\mathit{numSpils \leq pSortFactor^2}$.
In the opposite case, we must use a simulation-based approach in order
to calculate the number of spills merged during the intermediate merge
rounds as well as the total number of merge passes. \\

\noindent The first merge pass is also unique
because Hadoop will calculate the optimal number of spill files to merge
so that all other merge passes will merge exactly $\mathit{pSortFactor}$ files.

\noindent Since the Reduce task also contains a similar Merge Phase,
we define the following three methods to reuse later:

\beq\mathit{
calcNumSpillsFirstPass(N, F) =
  \begin{cases} 
   N & \hspace{-20mm} \text{, if } N \leq F \\
   F & \hspace{-20mm} \text{, if } (N - 1) \textit{ MOD } (F - 1) = 0 \\
   (N - 1) \textit{ MOD } (F - 1) + 1 & \text{, otherwise }
  \end{cases}
}\eeq

\begin{align}
\mathit{calcNumSpillsIntermMerge(N, F)} &=
  \mathit{
  \begin{cases} 
   0 & \text{, if } N \leq F \\
   P + \left\lfloor \frac{N - P}{F} \right\rfloor * F & \text{, if } N \leq F^2
  \end{cases}
  } \nonumber \\
  & \hspace{8mm}\mathit{, where\ P = calcNumSpillsFirstPass(N, F)}
\end{align}
\begin{align}
\mathit{calcNumSpillsFinalMerge(N, F)} &=
  \mathit{
  \begin{cases} 
   N & \text{, if } N \leq F \\
   1 + \left\lfloor \frac{N - P}{F} \right\rfloor + (N - S) & \text{, if } N \leq F^2
  \end{cases}
  } \nonumber \\
  & \hspace{8mm}\mathit{, where\ P = calcNumSpillsFirstPass(N, F)} \nonumber \\
  & \hspace{8mm}\mathit{, where\ S = calcNumSpillsIntermMerge(N, F)}
\end{align}

\noindent The number of spills read during the first merge pass is:

\beq\mathit{ numSpillsFirstPass = calcNumSpillsFirstPass(numSpills, pSortFactor) }\eeq

\noindent The number of spills read during the intermediate merging is:

\beq\mathit{numSpillsIntermMerge = calcNumSpillsIntermMerge(numSpills, pSortFactor) }\eeq

\noindent The total number of merge passes will be:

\beq\mathit{
numMergePasses = 
  \begin{cases} 
   0 & \text{, if } numSpills = 1 \\
   1 & \text{, if } numSpills \leq pSortFactor \\
   2 + \left\lfloor \frac{numSpills - numSpillsFirstPass}{pSortFactor} \right\rfloor
     & \text{, if } numSpills \leq pSortFactor^2
  \end{cases}
}\eeq

\noindent The number of spill files for the final merge round is
(first pass + intermediate passes + remaining file spills):

\beq\mathit{numSpillsFinalMerge = calcNumSpillsFinalMerge(numSpills, pSortFactor) }\eeq

\noindent The total number of records spilled is:

\beq\mathit{
numRecSpilled = spillFilePairs \times [numSpills + numSpillsIntermMerge
		+ numSpills \times sCombinePairsSel]
}\eeq

\noindent The final map output size and number of K-V pairs are:
\begin{align}
\mathit{useCombInMerge} &= \mathit{(numSpills > 1) \text{\ AND\ } (pUseCombine)} \nonumber \\
	& \ \ \text{\ AND\ } \mathit{(numSpillsFinalMerge \geq pNumSpillsForComb)}
\end{align}
\begin{align}
\mathit{intermDataSize} &= \mathit{numSpills \times spillFileSize}  \nonumber \\
& \ \ \ \times
   \begin{cases} 
   sCombineSizeSel & \text{if} \ useCombInMerge \\
   1 	& \text{otherwise}
   \end{cases}
\end{align}
\begin{align}
\mathit{intermDataPairs} &= \mathit{numSpills \times spillFilePairs}  \nonumber \\
& \ \ \times
   \begin{cases} 
   \mathit{sCombinePairsSel} & \text{if} \ \mathit{useCombInMerge} \\
   1 	& \text{otherwise}
   \end{cases}
\end{align}

\noindent  The costs of this phase are:
\begin{align}
\mathit{IOCost_{Merge}} &= & \nonumber \\
	&\hspace{-15mm} \ \mathit{2 \times numSpillsIntermMerge \times spillFileSize \times cLocalIOCost}
		& \hspace{-8mm}\text{// interm merges} \nonumber \\
	&\hspace{-15mm} \mathit{+ numSpills \times spillFileSize \times cLocalIOCost}
		& \hspace{-6mm}\text{// read final merge}\nonumber \\
	&\hspace{-15mm} \mathit{+ intermDataSize \times cLocalIOCost}
		& \hspace{-5mm}\text{// write final merge}
\end{align}
\begin{align}
\mathit{CPUCost_{Merge}} &= \nonumber \\
	&\hspace{-10mm} \mathit{numSpillsIntermMerge \times} \nonumber \\
    &\hspace{-5mm} [\ \mathit{spillFileSize\times cIntermUncomprCPUCost} \nonumber \\
	&\hspace{-5mm} \ \mathit{+ spillFilePairs \times cMergeCPUCost} \nonumber \\
    &\hspace{-5mm} \ \mathit{+ \frac{spillFileSize}{sIntermCompressRatio} 
				\times cIntermComprCPUCost\ ]} \nonumber \\
	&\hspace{-10mm} \mathit{+ numSpills \times} \nonumber \\
    &\hspace{-5mm} [\ \mathit{spillFileSize\times cIntermUncomprCPUCost} \nonumber \\
	&\hspace{-5mm} \ \mathit{+ spillFilePairs \times cMergeCPUCost} \nonumber \\
    &\hspace{-5mm} \ \mathit{+ spillFilePairs \times cCombineCPUCost}\ ] \nonumber \\
	&\hspace{-10mm}  \mathit{+ \frac{intermDataSize}{sIntermCompressRatio} 
				\times cIntermComprCPUCost}
\end{align}

\subsection{Modeling the Overall Map Task}
\label{sec:map-overall-task}

\noindent The above models correspond to the execution of a single map task.
The overall costs for a single map task are:

\beq\mathit{
IOCost_{Map} = 
  \begin{cases} 
   IOCost_{Read} + IOCost_{MapWrite}				& \text{if} \ pNumReducers = 0 \\
   IOCost_{Read} + IOCost_{Spill} + IOCost_{Merge} 	& \text{if} \ pNumReducers > 0
  \end{cases}
}\eeq

\beq\mathit{
CPUCost_{Map} = 
  \begin{cases} 
   CPUCost_{Read} + CPUCost_{MapWrite}					& \text{if} \ pNumReducers = 0 \\
   CPUCost_{Read} + CPUCost_{Spill} + CPUCost_{Merge} 	& \text{if} \ pNumReducers > 0
  \end{cases}
}\eeq

\section{Performance Models for the Reduce Task Phases}
\label{sec:reduce-phases}

\noindent The Reduce Task is divided into four phases:
\begin{enumerate}
{
\item {\em Shuffle}: Copying the map output from the mapper nodes 
to a reducer's node and decompressing, if needed.
Partial merging may also occur during this phase.

\item {\em Merge}: Merging the sorted fragments from the different mappers
to form the input to the reduce function.

\item {\em Reduce}: Executing the user-provided reduce function.

\item {\em Write}: Writing the (compressed) output to HDFS.
}
\end{enumerate}

\subsection{Modeling the Shuffle Phase}
\label{sec:reduce-shuffle-phase}

\noindent The following discussion refers to the execution of a single
reduce task. In the Shuffle phase, the framework fetches the relevant map
output partition from each mapper (called {\em segment})
and copies it to the reducer's node.
If the map output is compressed, it will be uncompressed.
For each map segment that reaches the reduce side we have:

\beq\mathit{ segmentComprSize = \frac{intermDataSize}{pNumReducers} }\eeq
\beq\mathit{ segmentUncomprSize = \frac{segmentComprSize}{sIntermCompressRatio} }\eeq
\beq\mathit{ segmentPairs = \frac{intermDataPairs}{pNumReducers} }\eeq

\noindent where $\mathit{intermDataSize}$ and $\mathit{intermDataPairs}$ are the size and number
of pairs produced as intermediate output by a single mapper
(see Section \ref{sec:map-merge-phase}). \\

\noindent The data fetched to a single reducer will be:

\beq\mathit{ totalShuffleSize = pNumMappers * segmentComprSize }\eeq
\beq\mathit{ totalShufflePairs = pNumMappers * segmentPairs }\eeq

\noindent As the data is copied to the reducer, they are placed in the 
shuffle buffer in memory with size:

\beq\mathit{ shuffleBufferSize = pShuffleInBufPerc \times pTaskMem }\eeq

\noindent When the in-memory buffer reaches a threshold size or the number
of segments becomes greater than the $\mathit{pInMemMergeThr}$, the segments
are merged and spilled to disk creating a new local file (called $\mathit{shuffleFile}$).
The merge size threshold is:

\beq\mathit{ mergeSizeThr = pShuffleMergePerc \times shuffleBufferSize }\eeq

\noindent However, when the segment size is greater that $25\%$ of 
the $\mathit{shuffleBufferSize}$, the segment will go straight to disk instead of
passing through memory (hence, no in-memory merging will occur). \\

\noindent {\bf Case 1:} $\mathit{segmentUncomprSize < 0.25 \times shuffleBufferSize}$

\beq\mathit{ numSegInShuffleFile = \frac{mergeSizeThr}{segmentUncomprSize} }\eeq
\begin{align}
	& \text{If } \mathit{(\lceil numSegInShuffleFile \rceil 
		\times segmentUncomprSize \leq shuffleBufferSize)} \nonumber \\
	& \ \ \ \mathit{numSegInShuffleFile = \lceil numSegInShuffleFile \rceil} \nonumber \\
	& \text{else} \nonumber \\
	& \ \ \ \mathit{numSegInShuffleFile = \lfloor numSegInShuffleFile \rfloor} \nonumber
\end{align}
\begin{align}
	& \text{If } \mathit{(numSegInShuffleFile > pInMemMergeThr)} \nonumber \\
	& \ \ \ \mathit{numSegInShuffleFile = pInMemMergeThr}
\end{align}

\noindent A shuffle file is the merging on $\mathit{numSegInShuffleFile}$ segments.
If a combine function is specified, then it is applied during this
merging. Note that if $\mathit{numSegInShuffleFile > numMappers}$, then merging
will not happen.

\begin{align}
\mathit{shuffleFileSize} &= \nonumber \\ 
	&\hspace{-15mm} \mathit{numSegInShuffleFile \times segmentComprSize \times sCombineSizeSel}
\end{align}
\begin{align}
\mathit{shuffleFilePairs} &= \nonumber \\ 
	&\hspace{-15mm} \mathit{numSegInShuffleFile \times segmentPairs \times sCombinePairsSel}
\end{align}
\beq\mathit{ numShuffleFiles = 
	\left\lfloor \frac{pNumMappers}{numSegInShuffleFile} \right\rfloor }\eeq

\noindent At the end of the merging, some segments might remain in memory.

\beq\mathit{ numSegmentsInMem = pNumMappers \text{ MOD } numSegInShuffleFile }\eeq

\noindent {\bf Case 2:} $\mathit{segmentUncomprSize \geq 0.25 \times shuffleBufferSize}$

\beq\mathit{ numSegInShuffleFile = 1 }\eeq
\beq\mathit{ shuffleFileSize = segmentComprSize }\eeq
\beq\mathit{ shuffleFilePairs = segmentPairs }\eeq
\beq\mathit{ numShuffleFiles = pNumMappers }\eeq
\beq\mathit{ numSegmentsInMem = 0 }\eeq

\noindent Either case will create a set of shuffle files on disk.
When the number of shuffle files on disk increases above a
certain threshold ($\mathit{2 \times pSortFactor - 1}$), a new merge thread is
triggered and $\mathit{pSortFactor}$ shuffle files are merged into a new larger
sorted one. The Combiner is not used during this disk merging.
The total number of such merges are:

\beq\mathit{
numShuffleMerges =
  \begin{cases} 
    0 \text{\hspace{20mm}, if } numShuffleFiles < 2 \times pSortFactor - 1 &  \\
    \left\lfloor \frac{numShuffleFiles - 2 \times pSortFactor + 1}{pSortFactor} 
	\right\rfloor + 1 \text{, otherwise} &
  \end{cases}
}\eeq

\noindent At the end of the Shuffle phase, a set of merged and unmerged
shuffle files will exist on disk.

\beq\mathit{ numMergShufFiles = numShuffleMerges }\eeq
\beq\mathit{ mergShufFileSize = pSortFactor \times shuffleFileSize }\eeq
\beq\mathit{ mergShufFilePairs = pSortFactor \times shuffleFilePairs }\eeq

\beq\mathit{ numUnmergShufFiles = numShuffleFiles - pSortFactor \times numShuffleMerges }\eeq
\beq\mathit{ unmergShufFileSize = shuffleFileSize }\eeq
\beq\mathit{ unmergShufFilePairs = shuffleFilePairs }\eeq

\noindent The cost of the Shuffling phase is:

\begin{align}
\mathit{IOCost_{Shuffle}}
	&= \mathit{numShuffleFiles \times shuffleFileSize \times cLocalIOCost} \nonumber \\
	& \hspace{-5mm} \mathit{+ numMergShufFiles \times mergShufFileSize 
				\times 2 \times cLocalIOCost}
\end{align}
\begin{align}
CPUCost_{Shuffle} &= \nonumber \\
	&\hspace{-25mm} \mathit{[\ totalShuffleSize \times cIntermUncomprCPUCost} \nonumber \\
	&\hspace{-20mm} \mathit{+ numShuffleFiles \times shuffleFilePairs 
						\times cMergeCPUCost} \nonumber \\
	&\hspace{-20mm} \mathit{+ numShuffleFiles \times shuffleFilePairs 
						\times cCombineCPUCost} \nonumber \\
	&\hspace{-20mm} \mathit{+ numShuffleFiles \times 
					\frac{shuffleFileSize}{sIntermCompressRatio} 
					\times cIntermComprCPUCost} \nonumber \\
	&\hspace{-25mm} \mathit{\ ] \times I(segmentUncomprSize < 0.25 
					\times shuffleBufferSize)} \nonumber \\
	&\hspace{-25mm} \mathit{+ numMergShufFiles \times mergShufFileSize 
					\times cIntermUncomprCPUCost} \nonumber \\
	&\hspace{-25mm} \mathit{+ numMergShufFiles \times mergShufFilePairs 
					\times cMergeCPUCost} \nonumber \\
	&\hspace{-25mm}	\mathit{+ numMergShufFiles \times 
					\frac{mergShufFileSize}{sIntermCompressRatio} 
					\times cIntermComprCPUCost}
\end{align}

\subsection{Modeling the Merge Phase}
\label{sec:reduce-merge-phase}

\noindent After all the map outputs have been successful copied in memory
and/or on disk, the sorting/merging phase begins. This phase will
merge all data into a single stream that is fed to the reducer.
Similar to the Map Merge phase (see Section \ref{sec:map-merge-phase}),
this phase may occur it multiple rounds, but during the final merging,
instead of creating a single output file, it will send the data
directly to the reducer. \\

\noindent The shuffle phase produced a set of merged and unmerged shuffle
files on disk, and perhaps a set of segments in memory.
The merging is done in three steps. \\

\noindent {\bf Step 1:} Some segments might be evicted from memory 
and merged into a single shuffle file to satisfy the
memory constraint enforced by $\mathit{pReducerInBufPerc}$. (This parameter
specifies the amount of memory allowed to be occupied by segments
before the reducer begins.)

\beq\mathit{ maxSegmentBuffer = pReducerInBufPerc \times pTaskMem }\eeq
\beq\mathit{ currSegmentBuffer = numSegmentsInMem \times segmentUncomprSize }\eeq
\begin{align}
	& \text{If } \mathit{(currSegmentBuffer > maxSegmentBuffer)} \nonumber \\
	& \ \ \ \mathit{ numSegmentsEvicted = 
		\left\lceil \frac{currSegmentBuffer - maxSegmentBuffer}
		{segmentUncomprSize} \right\rceil} \nonumber \\
	& \text{else} \nonumber \\
	& \ \ \ \mathit{numSegmentsEvicted = 0}
\end{align}
\beq\mathit{ numSegmentsRemainMem = numSegmentsInMem - numSegmentsEvicted }\eeq

\noindent The above merging will only occur if the number of existing 
shuffle files on disk are less than the $\mathit{pSortFactor}$. If not, then
the shuffle files would have to be merged, and the in-memory segments
that are supposed to be evicted are left to be merge with the shuffle
files on disk.

\beq\mathit{ numFilesOnDisk = numMergShufFiles + numUnmergShufFiles }\eeq
\begin{align}
	& \text{If } \mathit{(numFilesOnDisk < pSortFactor)} \nonumber \\
	& \ \ \ \mathit{numFilesFromMem = 1} \nonumber \\
	& \ \ \ \mathit{filesFromMemSize = numSegmentsEvicted \times segmentComprSize} \nonumber \\
	& \ \ \ \mathit{filesFromMemPairs = numSegmentsEvicted \times segmentPairs} \nonumber \\
	& \ \ \ \mathit{step1MergingSize = filesFromMemSize} \nonumber \\
	& \ \ \ \mathit{step1MergingPairs = filesFromMemPairs} \nonumber \\
	& \text{else} \nonumber \\
	& \ \ \ \mathit{numFilesFromMem = numSegmentsEvicted} \nonumber \\
	& \ \ \ \mathit{filesFromMemSize = segmentComprSize} \nonumber \\
	& \ \ \ \mathit{filesFromMemPairs = segmentPairs} \nonumber \\
	& \ \ \ \mathit{step1MergingSize = 0} \nonumber \\
	& \ \ \ \mathit{step1MergingPairs = 0}
\end{align}
\beq\mathit{ filesToMergeStep2 = numFilesOnDisk + numFilesFromMem }\eeq

\noindent {\bf Step 2:} Any files on disk will go through a merging phase
in multiple rounds (similar to the process in Section \ref{sec:map-merge-phase}.
This step will happen only if $\mathit{numFilesOnDisk > 0}$
(which implies $\mathit{filesToMergeStep2 > 0}$).
The number of intermediate reads (and writes) are:

\beq\mathit{ intermMergeReads = calcNumSpillsIntermMerge(filesToMergeStep2, pSortFactor) }\eeq

\noindent The main difference from Section \ref{sec:map-merge-phase}
is that the merged files have different sizes. We account for this by
attributing merging costs proportionally.

\begin{align}
\mathit{step2MergingSize} &= \mathit{\frac{intermMergeReads}{filesToMergeStep2} \times} \nonumber \\
	& \ \ \mathit{[\ numMergShufFiles \times mergShufFileSize} \nonumber \\
	& \ \ \mathit{+ numUnmergShufFiles \times unmergShufFileSize} \nonumber \\
	& \ \ \mathit{+ numFilesFromMem \times filesFromMemSize ]}
\end{align}
\begin{align}
\mathit{step2MergingPairs} &= \mathit{\frac{intermMergeReads}{filesToMergeStep2} \times} \nonumber \\
	& \ \ \mathit{[\ numMergShufFiles \times mergShufFilePairs} \nonumber \\
	& \ \ \mathit{+ numUnmergShufFiles \times unmergShufFilePairs} \nonumber \\
	& \ \ \mathit{+ numFilesFromMem \times filesFromMemPairs ]}
\end{align}

\beq\mathit{ filesRemainFromStep2 = calcNumSpillsFinalMerge(filesToMergeStep2, pSortFactor) }\eeq

\noindent {\bf Step 3:} All files on disk and in memory will go through
merging.

\beq\mathit{ filesToMergeStep3 = filesRemainFromStep2 + numSegmentsRemainMem }\eeq

The process is identical to step 2 above.

\beq\mathit{ intermMergeReads = calcNumSpillsIntermMerge(filesToMergeStep3, pSortFactor) }\eeq
\beq\mathit{ step3MergingSize = 
	\frac{intermMergeReads}{filesToMergeStep3} \times totalShuffleSize }\eeq

\beq\mathit{ step3MergingPairs = 
	\frac{intermMergeReads}{filesToMergeStep3} \times totalShufflePairs }\eeq
\beq\mathit{ filesRemainFromStep3 = calcNumSpillsFinalMerge(filesToMergeStep3, pSortFactor) }\eeq
\beq\mathit{ totalMergingSize = step1MergingSize + step2MergingSize + step3MergingSize }\eeq

\noindent The cost of the Sorting phase is:

\begin{align}
\mathit{IOCost_{Sort} = totalMergingSize \times cLocalIOCost}
\end{align}
\begin{align}
\mathit{CPUCost_{Sort}} &= \nonumber \\
	&\hspace{-10mm} \mathit{totalMergingSize \times cMergeCPUCost} \nonumber \\
	&\hspace{-10mm} \mathit{[\frac{totalMergingSize}{sIntermCompressRatio} ] 
				\times cIntermComprCPUCost} \nonumber \\
	&\hspace{-10mm} \mathit{[step2MergingSize + step3MergingSize] \times cIntermUnomprCPUCost}
\end{align}

\subsection{Modeling the Reduce and Write Phases}
\label{sec:reduce-write-phase}

\noindent Finally, the user-provided reduce function will be executed
and the output will be written to HDFS.

\begin{align}
\mathit{inReduceSize} &
	= \mathit{\frac{numShuffleFiles \times shuffleFileSize}{sIntermCompressRatio}} \nonumber \\
	& \ \ \mathit{+ \frac{numSegmentsInMem \times segmentComprSize}{sIntermCompressRatio}}
\end{align}
\begin{align}
\mathit{inReducePairs} &
	= \mathit{numShuffleFiles \times shuffleFilePairs} \nonumber \\
	& \ \ \mathit{+ numSegmentsInMem \times segmentComprPairs}
\end{align}

\beq\mathit{ outReduceSize = inReduceSize \times sReduceSizeSel }\eeq
\beq\mathit{ outReducePairs = inReducePairs \times sReducePairsSel }\eeq

\noindent The input to the reduce function resides in memory and/or in
the shuffle files produced by the Shuffling and Sorting phases.

\begin{align}
\mathit{inRedSizeDiskSize} &= \mathit{numMergShufFiles \times mergShufFileSize} \nonumber \\
	& \ \ \mathit{+ numUnmergShufFiles \times unmergShufFileSize} \nonumber \\
	& \ \ \mathit{+ numFilesFromMem \times filesFromMemSize}
\end{align}

\noindent The cost of the Write phase is:

\begin{align}
\mathit{IOCost_{Write}} &= \mathit{inRedSizeDiskSize \times cLocalIOCost} \nonumber \\
	& \ \  \mathit{+ outReduceSize \times sOutCompressRatio \times cHdfsWriteCost}
\end{align}
\begin{align}
\mathit{CPUCost_{Write}} &= \mathit{inReducePairs \times cReduceCPUCost} \nonumber \\
	& \ \ \mathit{+ inRedSizeDiskSize \times cIntermUncompCPUCost} \nonumber \\
	& \ \ \mathit{+ outReduceSize \times cOutComprCPUCost}
\end{align}

\subsection{Modeling the Overall Reduce Task}
\label{sec:reduce-overall-task}

\noindent The above models correspond to the execution of a single reduce task.
The overall costs for a single reduce task, excluding network transfers, are:

\beq\mathit{ IOCost_{Reduce} = 
   IOCost_{Shuffle} + IOCost_{Sort} + IOCost_{Write}
}\eeq
\beq\mathit{
CPUCost_{Reduce} = 
   CPUCost_{Shuffle} + CPUCost_{Sort} + CPUCost_{Write}
}\eeq

\section{Performance Models for the Network Transfer}
\label{sec:network-transfer}

\noindent During the shuffle phase, all the data produced by the map
tasks is copied over to the nodes running the reduce tasks (except for
the data that is local). The overall data transferred in the network is:

\beq\mathit{ netTransferSize = finalOutMapSize \times pNumMappers \times \frac{pNumNodes-1}{pNumNodes} }\eeq

\noindent where $\mathit{finalOutMapSize}$ is the size of data produced by
a single map tasks. \\

\noindent The overall cost for transferring data over the network is:

\beq\mathit{ NETCost_{Job} = netTransferSize \times networkCost }\eeq

\section{Performance Models for the Map-Reduce Job}
\label{sec:mr-job}

\noindent The MapReduce job consists of several map and reduce tasks executing
in parallel and in waves. There are two primary ways to estimating the
total costs of the job: (i) simulate the task execution
using a {\em Task Scheduler Simulator}, and (ii) calculate the expected
total costs analytically. \\

Simulation involves scheduling and simulating
the execution of individual tasks on a virtual Cluster. The cost for
each task is calculated using the proposed performance models. \\

The second approach involves using the following analytical costs:

\beq\mathit{ IOCost_{AllMaps} = 
	\frac{pNumMappers \times IOCost_{Map}}{pNumNodes \times pMaxMapsPerNode}
}\eeq

\beq\mathit{ CPUCost_{AllMaps} = 
	\frac{pNumMappers \times CPUCost_{Map}}{pNumNodes \times pMaxMapsPerNode}
}\eeq

\beq\mathit{ IOCost_{AllReducers} = 
	\frac{pNumReducers \times IOCost_{Reduce}}{pNumNodes \times pMaxRedPerNode}
}\eeq

\beq\mathit{ CPUCost_{AllReducers} = 
	\frac{pNumReducers \times CPUCost_{Reduce}}{pNumNodes \times pMaxRedPerNode}
}\eeq

\noindent The overall job cost is simply the sum of the costs from all the
map and the reduce tasks.

\beq\mathit{
IOCost_{Job} = 
  \begin{cases} 
   IOCost_{AllMaps} 						& \text{if} \ pNumReducers = 0 \\
   IOCost_{AllMaps} + IOCost_{AllReducers}	& \text{if} \ pNumReducers > 0
  \end{cases}
}\eeq

\beq\mathit{
CPUCost_{Job} = 
  \begin{cases} 
   CPUCost_{AllMaps} 							& \text{if} \ pNumReducers = 0 \\
   CPUCost_{AllMaps} + CPUCost_{AllReducers}	& \text{if} \ pNumReducers > 0
  \end{cases}
}\eeq

\noindent With appropriate system parameters that allow for equal comparisons
among the I/O, CPU, and network costs, the overall cost is:

\beq\mathit{ Cost_{Job} = IOCost_{Job} + CPUCost_{Job} + NETCost_{Job} }\eeq

\end{document}